\documentclass[preprint,authoryear,12pt]{elsarticle2}
\usepackage{epsfig}
\usepackage{graphicx}
\usepackage{amssymb,amsmath}

\def\figfigeps#1#2#3{\centerline{\epsfig{figure=#2.eps,height=#1}}%
\caption{\small #3}\label{fig:#2}}

\newcommand{\R}{\mathbb{R}}
\newcommand{\x}{\mathbf{x}}
\newcommand{\y}{\mathbf{y}}
\newcommand{\z}{\mathbf{z}}
\journal{Icarus}
\begin{document}

\title{Identification of known objects in solar system surveys}

\author[pi]{Andrea Milani}
\ead{milani@dm.unipi.it}
\author[bg]{Zoran Kne\v zevi\'c}

\author[spacedys,pi]{Davide Farnocchia}
\author[spacedys]{Fabrizio Bernardi}
\author[ifa]{Robert Jedicke}
\author[ifa]{Larry Denneau}

\author[ifa]{Richard J. Wainscoat}
\author[ifa]{William Burgett}
\author[psi]{Tommy Grav}
\author[ifa]{Nick Kaiser}
\author[ifa]{Eugene Magnier}
\author[das]{Paul A. Price}

\address[pi]{Dipartimento di Matematica, Universit\`a di Pisa,
        Largo Pontecorvo 5,
        56127 Pisa, Italy}
\address[bg]{Astronomical Observatory, Volgina 7,
         11060 Belgrade 38, Serbia}
\address[spacedys]{SpaceDyS, Via Mario Giuntini 63, 56023 Cascina,
 Pisa, Italy}
\address[ifa]{Institute for Astronomy, University of Hawaii, 2680 Woodlawn Drive, Honolulu, HI, USA}
\address[psi]{Planetary Science Institute, 1700 East Fort Lowell, Suite 106, Tucson, AZ 85719}
\address[das]{Department of Astrophysical Sciences, Princeton University, Princeton, NJ 08544, USA}

\date{submitted, January 12, 2012; revised March 19, 2012}
\begin{abstract}

The discovery of new objects in modern wide-field asteroid and comet
surveys can be enhanced by first identifying observations belonging
to known solar system objects. The assignation of new observations
to a known object is an {\it attribution problem} that occurs when a
least squares orbit already exists for the object but a separate fit
is not possible to just the set of new observations.  In this work
we explore the strongly asymmetric attribution problem in which the
existing least squares orbit is very well constrained and the new
data are sparse.
We describe an attribution algorithm that introduces new quality
control metrics in the presence of strong biases in the astrometric
residuals. The main biases arise from the stellar catalogs used in the
reduction of asteroid observations and we show that a simple debiasing
with measured regional catalog biases significantly improves the
results.
We tested the attribution algorithm using data from the PS1 survey
that used the 2MASS star catalog for the astrometric reduction.  We
found small but statistically significant biases in the data of up to
0.1 arcsec that are relevant only when the observations reach the
level of accuracy made possible by instruments like PS1.
The false attribution rate was measured to be $< 1/1,000$ with a simple
additional condition that can reduce it to zero while the attribution
efficiency is consistent with 100\%.

\end{abstract}

\maketitle


\section{The problem}

When surveying the sky with the purpose of discovering new solar
system objects known asteroids and comets are also being identified. It
is desirable to separately treat the identification of the known
moving objects for four reasons: to avoid claiming as a new discovery
some well known object, to reduce the dataset while searching for new
discoveries, to improve the orbits of the known objects while ensuring
they are not contaminated by false associations and finally, for
statistical quality control of the astrometric data using the
residuals with the well known orbits.

The procedure of assigning new observations to known objects is a
special case of the class of \textit{identification problems}
\citep[Chap.~7]{orbdet}. The problem arises because the same object is
observed in short arcs separated by typically many years.  The goal is
to build the list of the observations belonging to the same physical
object without introducing any \textit{false} detections corresponding
to another moving object, image defect, statistical fluke, or
non-solar system object.

The \textit{attribution problem} occurs when a least squares orbit has
already been fit to a set of historical observations of an object and
we want to assign a new set of observations for which it is not
possible to independently fit a least squares orbit.\footnote{For a
  full description of the terminology see \citet{marsden85, ident1}
  and \citet[Chap. 7]{orbdet}.}  The methods for finding and
confirming attributions were described in \citet{ident4} and were
shown to be effective for both simulations and real data in
\citet{ons2} \citep[see also, e.g.,\ ][]{gra08,san11}. \citet{ons3}
then developed a high reliability statistical quality control
procedure to confirm the attributions.  The algorithms developed to
date have been successful when the new data sets contain less
information but are comparable in quantity and quality to the old one.

In this paper we are interested in the special case in which the
previously known asteroids have well constrained orbits while the new
data to be attributed are sparse.  There is typically just one
\textit{tracklet}, a very short arc of astrometric observations,
assembled by the observer using just a linear or a quadratic
fit to the astrometry as a function of time.  The challenge
is to account for the data asymmetry: a small number of new accurate
observations per object from the state of the art surveys to be
associated with the historical data set containing many low accuracy
observations per object.

The main difficulty of the asymmetric identification problem is that
both the existing and new data are biased by the astrometric catalogs
and a reliable astrometric error model (including rigorously estimated
RMS and correlations) is usually not available.  As a consequence, the
better the orbit is constrained by the existing data the worse is the
effect of the biases on the ephemerides predicted from the orbit and
its covariance matrix.  In the idealized case the existing data may
have a larger standard deviation but be unbiased (zero average
astrometric error) and the observational errors could be modeled by a
normal distribution with known RMS.

A quantitative and real example might be more convincing than a
theoretical argument.  The
\emph{AstDyS}\footnote{http://hamilton.dm.unipi.it/astdys,
  as of November 2011.}  online service tells us that at the current
date there are $68,571$ \emph{numbered} or \emph{multi-apparition}
asteroids ($76\%$ of those with apparent magnitude $<22$ and solar
elongation $>120^\circ$) with a formal RMS of the current ephemerides
prediction $\leq 0.3$~arcsec.  Thus, a bias in the astrometry of the
order of $0.5$ arcsec would result in a systematic error larger than
the estimated random error.  We show that the first of the next
generation surveys, Pan-STARRS1, \citep[PS1,\ ][]{hodapp2004}
generates astrometric data with accuracy of $0.1$ to $0.15$ arcsec ---
of such high quality that if fit to orbits computed with biased
historic data the two data sets would appear statistically
incompatible. Thus, it would be very difficult to validate the
accuracy of the new data unless debiased data were available (both
historic and new).

Our two-step solution is to (i) devise a new statistical quality
control procedure which is applied asymmetrically to the old and new
data  and (ii) apply a debiasing procedure
to the MPC-archived historical observations to remove the known
position-dependent bias due to regional systematic errors in the
astrometric star catalogs \citep{cbm10}.

We tested our new methods for asymmetric attribution using early
results from the PS1 telescope. The source orbits used were numbered
and multi-apparition asteroids. The purpose of the test was to
validate our asymmetric attribution method and measure the PS1
system's astrometric accuracy. In particular, we were interested in
constructing a rigorous PS1 astrometric error model taking into
account the correlations. as described 

The structure of this paper is as follows. The known algorithms for
attribution are summarized in Section~\ref{s:oldatt}. The new
algorithms to be applied specifically to the asymmetric case are
introduced in Section~\ref{s:knownatt}. In Section~\ref{s:debias} 
we show the importance of the removal of the star catalog biases.
In Section~\ref{s:panstarrs} we describe the results of two tests of
our methods, both using PS1 data, and a first attempt at building a
rigorous statistical error model for PS1 astrometry.
Section~\ref{s:acceff} gives the results of the tests for the accuracy
and efficiency of our algorithms using real data from the PS1
survey.  In Section~\ref{s:conclusions} we draw some conclusions and
indicate possible directions for future work.


\section{The attribution algorithms}
\label{s:oldatt}

An \textit{attribution problem} requires that there exist 1) a vector
$\x_1\in \R^6$ of orbital elements at epoch $t_1$ along with a
$6\times 6$ \textit{covariance matrix} $\Gamma_{\x_1}$ describing
their uncertainty in the linearized approximation and 2) a vector
$\y_o\in \R^s$ of \textit{observables} ($s<6$) at another epoch $t_2$.
Let $\y_c=F(\x_1,t_2)$ be the predicted set of observables at time
$t_2$ using the covariance $\Gamma_{\x_1}$ to \textit{propagate the
  covariance matrix to the space of the observables} by the linearized
formula
\[
\Gamma_{\y_c}=DF\, \Gamma_{\x_1} \, DF^T
\]
where $DF$ is the $s\times 6$ Jacobian matrix of $F$ computed at
$\x_1$.  Using $\Gamma_{\y_c}$ we can assess the likelihood of the
prediction $\y_c$ being compatible with the hypothesis that the
observation $\y_o$ belongs to the same object within the linear
approximation and assuming an unbiased Gaussian error model.

For reasons of efficiency in handling large numbers of both orbits and
tracklets this process should be implemented using a sequence of
filters as described below. Each step provides an increasing
likelihood of identification with an increasing computational effort
applied to a decreasing number of possible cases \citep{ident4}.

\subsection{Filter 1: Observed-Computed residuals on the celestial
  sphere}
\label{s:filter1}

The first test is to compare the new angular observations to the
prediction on the celestial sphere in right ascension and declination
$(\alpha, \delta)$.  The difference $\y_o-\y_c=\Delta\y=(\Delta\alpha,
\Delta\delta)\in \R^2$ is projected on the \textit{tangent plane to
  the celestial sphere}.  From the $2\times 2$ covariance matrix
$\Gamma_{\y_c}$ we compute the \textit{normal matrix}
$C_{\y_c}=\Gamma_{\y_c}^{-1}$ and the {confidence ellipse}
\[
Z_{\y_c}(\sigma_c)=\{\Delta\y | \Delta \y^T \, C_{\y_c}\,\Delta\y \leq
\sigma_c^2\}\ .
\]
The observation $\y_o$ has its own uncertainty given by the $2\times
2$ covariance matrix $\Gamma_{\y_o}$ and the normal matrix
$C_{\y_o}=\Gamma_{\y_o}^{-1}$ yielding a second confidence ellipse
$Z_{\y_o}(\sigma_o)$ for the observed angles.  In most cases we can
assume the observation error is isotropic because the astrometric
reduction process typically makes no distinction between the two
directions. I.e., $Z_{\y_o}$ can be assumed to be just a
\textit{circle} in the $(\cos\delta \Delta\alpha, \Delta\delta)$
coordinates.

This filtering step requires that the two ellipses intersect for
reasonable values of the confidence parameters $\sigma_c$ and
$\sigma_o$.  Since this filter may have to be applied for each orbit
$\x_1$ to a large set of new observations $\y_o$ it is often
convenient to resort to a simplified test using only the larger of
the two ellipses. E.g., $\Delta\y\in Z_{\y_c}(\sigma_c)$ when the
uncertainty of the prediction is larger and $\Delta\y\in
Z_{\y_o}(\sigma_o)$ when the observational errors are larger.

When the orbit $\x_1$ is already well determined (as assumed in this
work) the two uncertainties can be comparable. A fast algorithm to
handle this case is to select a control $\sigma_c$ for
$Z_{\y_c}(\sigma_c)$ and a circular radius $\sigma_o$ for
$Z_{\y_o}(\sigma_o)$ and then compute the inequality defining an
ellipse with the semiaxes of $Z_{\y_c}(\sigma_c)$ increased in length
by $\sigma_o$ in the $(\cos\delta \Delta\alpha, \Delta\delta)$
coordinates.

This algorithm can be applied either to a single observation belonging
to the set of observations for which attribution is attempted (the
tracklet) or, better yet, to an \emph{average} observation obtained
from all those forming a tracklet because the averaging removes some
of the random error.

\subsection{Filter 2: Attributables and attribution penalties}
\label{s:filter2}

Tracklets containing observations spanning a short time (typically 15
min to 2 hours) can generally be compressed into an
\textit{attributable} vector $\y=(\alpha, \delta, \dot\alpha,
\dot\delta)\in\R^4$ containing both the angular position and angular
motion vector. In the special case of objects that are very close to
the Earth there is additional curvature information that may further
constrain the range and range-rate \citep{ons3}. The attributable
vector is obtained by a linear fit to the observations in the tracklet
resulting in the observable $\z_o\in \R^4$ at the average time $t_2$
with a $4\times 4$ normal matrix $C_{\z_o}$.

The ``distance'' between the observed and predicted attributables is
given by a metrics that takes into account the
uncertainty of both data points using their respective
normal matrices.  The predicted attributable vector at time $t_2$ from
the state $(\x_1, t_1)$ is nominally $\z_c\in \R^4$ with a $4\times 4$
normal matrix $C_{\z_c}$.  Its compatibility with the observed
attributable vector can be described by the \textit{attribution
  penalty} $K_4$ \citep[Chap. 7][]{orbdet,ident4} defined by
\begin{eqnarray*}
C_0&=&C_{\z_c}+C_{\z_o}\ \ ; \ \ \Gamma_0=C_0^{-1}\nonumber\\
K_4&=& (\z_o-\z_c)^T\, \left [C_{\z_c}
-C_{\z_c}\,\Gamma_0\,C_{\z_c}\right]\,(\z_o-\z_c)\
\end{eqnarray*}
\noindent that corresponds geometrically to testing the intersection
of the confidence ellipsoids with $\sigma_4=\sqrt{K_4}$ playing the
role of the confidence parameter.

We can discard attribution candidates when the attribution penalty
exceeds a reasonable value of $\sigma_4^2$.  If the predictions were
perfectly linear and an exact Gaussian error model was available for
the observations then a $\chi^2$ table could be used to select
$\sigma_4^2$ but these conditions are never satisfied with real data.
Thus, the parameter should be determined empirically with simulations
and/or tests with real data.

\subsection{Filter 3: Differential corrections and Quality control
  metrics}
\label{s:filter3}

Attributions that pass the second filter need to be confirmed by a
least squares fit with all the available observations. The fit must
converge and yield residuals with good statistical properties
compatible with what we know about the observational errors. The most
common control is the RMS of the residuals weighted with their
accuracy.

The attribution reliability can be increased with \textit{statistical
  quality controls} based upon more than one parameter that also
capture information based on systematic signals remaining in the
residuals.  We normally use the following 10 metrics:
\begin{itemize}
\item $RMS$ root mean square of the astrometric
  residuals\footnote{Residuals in $\alpha$ are multiplied by
    $\cos(\delta)$ before computing all the metrics.},
\item $BIAS_\alpha, BIAS_\delta$ bias, i.e., average of the
  astrometric residuals,
\item $SPAN_\alpha, SPAN_\delta$ first derivatives of the astrometric
  residuals,
\item $CURV_\alpha, CURV_\delta$ second derivatives of the astrometric
  residuals,
\item $ZSIGN_\alpha, ZSIGN_\delta$ third derivatives of the
  astrometric residuals,
\item $RMSH$ RMS of photometric residuals in magnitudes.
\end{itemize}
To use the same controls for both historic and new data (that are
typically of different accuracy) we normalize the bias and derivatives
of the residuals by fitting them to a polynomial of degree 3 and
dividing the coefficients by their standard deviation obtained from
the covariance matrix of the fit.

This filter has been shown to be effective in removing false
identifications \citep{ons3} using Pan-STARRS survey simulations.  The
advantage of using simulations is that it is then possible to use the
\textit{ground truth}, i.e., the {\it a priori} knowledge of the
correspondence between objects and their observations, to verify the
identifications. The control parameters can then be adjusted to obtain
the desired balance between efficiency and accuracy. The values of the
controls selected in this way are close to 1 as expected from Gaussian
statistics \citep[][, Sec. 6]{ons3} because the orbits are already
well constrained.  Note that the survey which was simulated does not
correspond exactly to the realized PS1 survey.  What matters for our
argument is that, since the false identifications are proportional to
the square of the number of tracklets, it was a pseudo-realistic large
scale simulation with a huge number of `known objects' and detected
tracklets.

\section{Asymmetric attribution quality control}
\label{s:knownatt}

The algorithms to be used to propose attributions in a strongly
asymmetric situation (such as high-accuracy numbered asteroid orbits
with new un-attributed tracklets) can be more or less as described
above, but the methods to confirm the attributions require additional
controls. The main reason is the presence of biases in the astrometric
errors.

\begin{figure}[t]
  \figfigeps{10 cm}{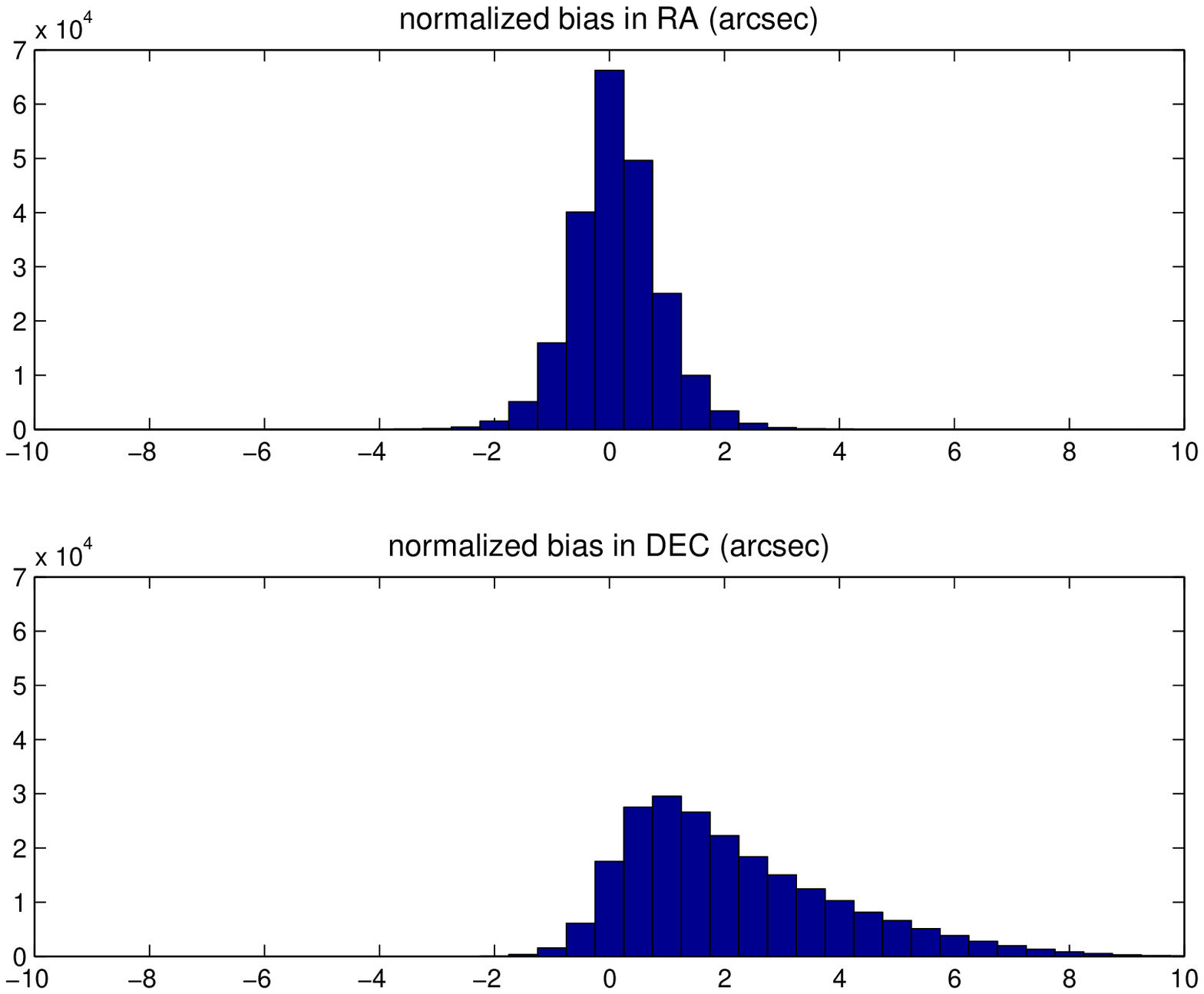}{Histogram of normalized biases
    in the astrometric residuals for all numbered asteroids from
    AstDyS (April 2010). Top: residuals in right ascension (multiplied
    by $\cos(\delta)$). Bottom: residuals in declination.}
\end{figure}

Figure~\ref{fig:bias_astdys_apr10} shows that the normalized biases of
astrometric residuals for the over-determined orbits of numbered
asteroids are strikingly different from a zero-mean Gaussian. The
normalized declination bias is qualitatively different from a Gaussian
with a mean of $\sim 2.17$ and a standard deviation of $\sim 1.86$.
In right ascension the shape is not too different from a normal
distribution but with a non-zero mean of $\sim 0.12$ and standard
deviation of $\sim 0.77$.  The span, curvature and Z-sign have similar
non-Gaussian distributions although the declination bias is the worst
case.  The causes of this behavior and the methods to mitigate their
effects are discussed in Section~\ref{s:debias}.

We need to take into account that the values of the quality metrics
are already high even before the attribution when attributing new
observations to objects with strongly over-determined orbits. Thus, we
cannot use small values of the control parameters because it would
generate the paradoxical result that the already computed orbit should
be discarded before adding new observations. Even if the new
observations improve the orbit determination because they are of
better quality and/or extend the observational time span, the
systematic error signatures contained in the residuals of the existing
orbit are not going to be removed. On the other hand, if we use high
values of the controls we may fail to reject some false attributions.

Our solution is to take into account the values of the metrics {\it
  and} the amount they change as a result of the proposed
attribution. E.g., for the $RMS$ metrics we accept an attribution only
if the increase resulting from the attribution is small (we used
$<0.15$ in Table~\ref{tab:deb_panst}). For the other 8 metrics that
are based upon astrometric residuals we set an upper limit to the
increase in the absolute value (we used $<2$ in
Table~\ref{tab:deb_panst}).

\subsection{Quality control for new observations}
\label{s:qc2}

The new observations may have little effect on the statistical
properties of the complete set of residuals because the tracklets
contain typically only $2$ to $8$ observations while the existing data
set typically contains tens or even hundreds of observations, i.e.,
the residuals may be neither significantly better nor significantly
worse.  Thus, we need to separately consider the residuals of the
attributed observations for which we require the $RMS$, $BIAS$ and
$SPAN$ (typical tracklets have no significant curvature and even less
Z-sign).  For the tests reported in Table~\ref{tab:deb_panst} we have
used the controls $RMS <3,\;|BIAS| <3,\; |SPAN|<3$. These values were
chosen empirically based on our experience with both real data and
simulations as reported in \citet{ons2}. Our experience suggests that
it is not possible to use theoretically motivated control values from
e.g., a $\chi^2$ table although they must be of the theoretical order.

The attribution procedure is recursive with tracklets added one by one
proceeding in order from the most to least likely as measured by the
penalty $K_4$. When an attribution has passed all the quality controls
and a new orbit has been fit to all the observations including the new
tracklet, the new orbit becomes the object's reference orbit. Then,
the other tracklets proposed for attribution from filter 2 are passed
to filter 3 and quality control, and so on.

This procedure is robust but we cannot claim that it is perfectly
reliable. Due to the stochastic nature of the observational errors any
attribution may be wrong; even those that result in the determination
of a numbered asteroid orbit, although the probability of this
happening has to be very small.

Despite the probability being small, when there are a large number of
trials the odds of finding low probability events approaches unity.
Our processing attributed a 4-observation PS1 tracklet from the year
2010 to the numbered asteroid (229833) 2009 BQ$_{25}$; however, the
fit to all the available data required discarding 6 observations that
were designated 2000~KU$_{51}$ from the apparition in the year
2000. In this case the only way to decide which attribution to
(229833) is correct might be to reexamine the original observations.

\section{Data debiasing for historic data}
\label{s:debias}

\begin{figure}[h]
  \figfigeps{10 cm}{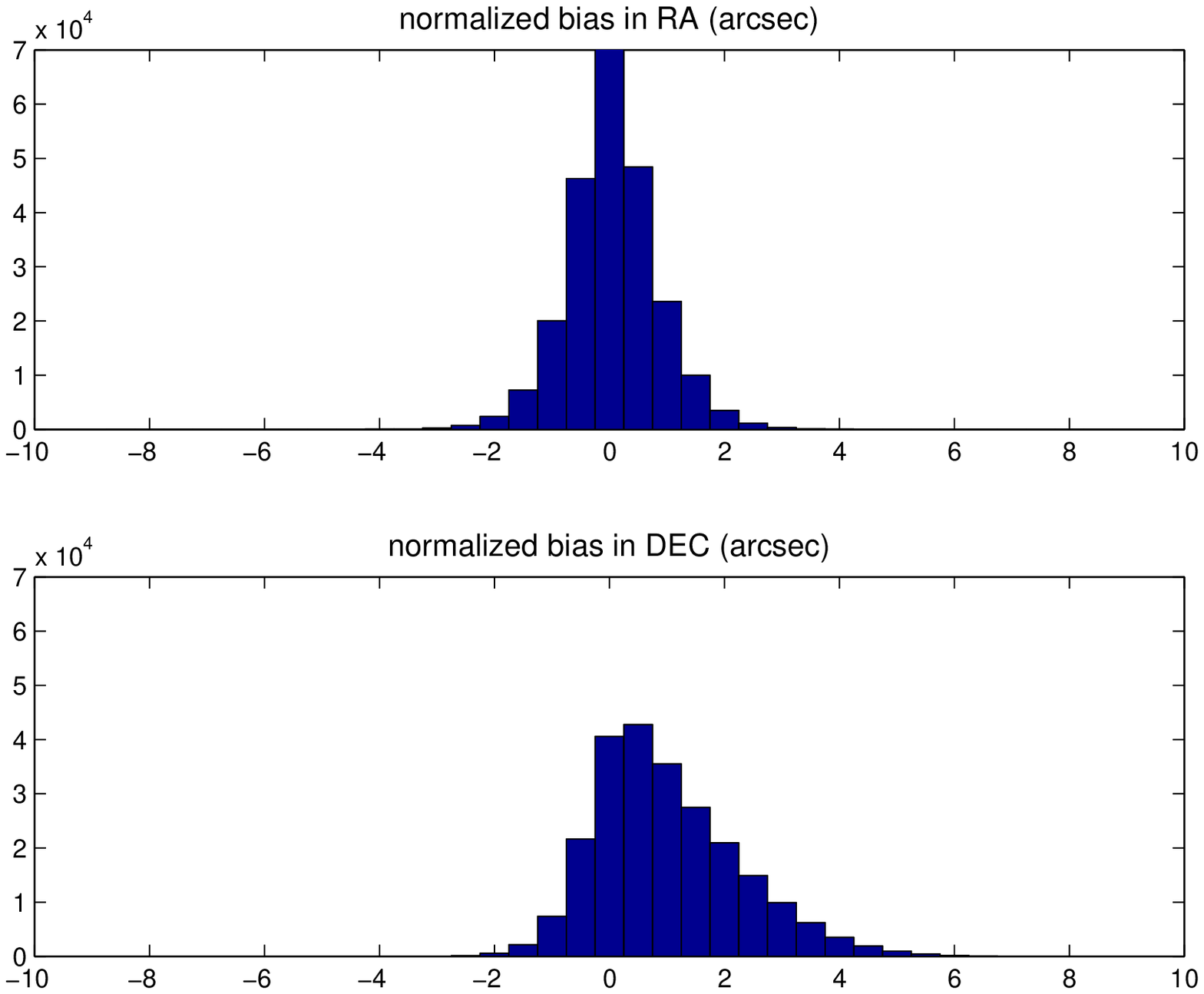}{Histogram of normalized
    astrometric residual biases for all numbered asteroids as in
    Figure~\ref{fig:bias_astdys_apr10} but after removing systematic
    star catalog errors from all the observations.}
\end{figure}

The most important source of systematic errors in observations of
solar system objects is in the star catalog used for astrometric
reduction.  \citet{cbm10} suggested that the errors could be mitigated
by debiasing the astrometric asteroid data with the measured regional
catalog biases. The biases can be computed as the average difference
in position between the stars in one catalog with respect to another
more accurate catalog. \citet{cbm10} used the 2MASS star catalog
\citep{2MASS} as their reference because it is of good accuracy and
covers the entire sky with a sufficient number of stars per unit area.
The catalog used for the astrometric reduction of each asteroid
observation is available for $92.8\%$ of the existing CCD data.

We have implemented an error model consistent with \citet{cbm10} by 1)
debiasing the observations by subtracting their calculated biases and
2) assigning a per-observation weight that is inversely proportional
to the debiased RMS residuals given in \citet[Table 6]{cbm10} (for the
same observatory and the same catalog when known). More precisely, the
weight was $1/(2\cdot RMS)$ when known but set to $1/1.5$
arcsec$^{-1}$ for data after 1950, for observations performed by
photographic techniques, and for CCD observations that do not include
star catalog information.

Figure~\ref{fig:bias_debiased_apr10} shows that the bias distribution
is much less asymmetric after implementation of the error model. The
mean of the normalized biases is now $\sim 1.00$ in declination
(standard deviation $\sim 1.26$) and $\sim 0.05$ in right ascension
(standard deviation $\sim 0.80$). The debiased set of declinations is
still biased but improves by a factor $\sim 2$ and the debiased right
ascensions has biases at the level of the quality of the best catalogs
and is as good as it can currently be.  Thus, the observation
debiasing and weighting improves the performance of the asymmetric
attribution algorithms but highlights that improvements in the star
catalogs are still necessary to take advantage of modern high-accuracy
asteroid astrometry.

\begin{figure}[h]
  \centerline{\includegraphics[width=12cm,height=11cm]{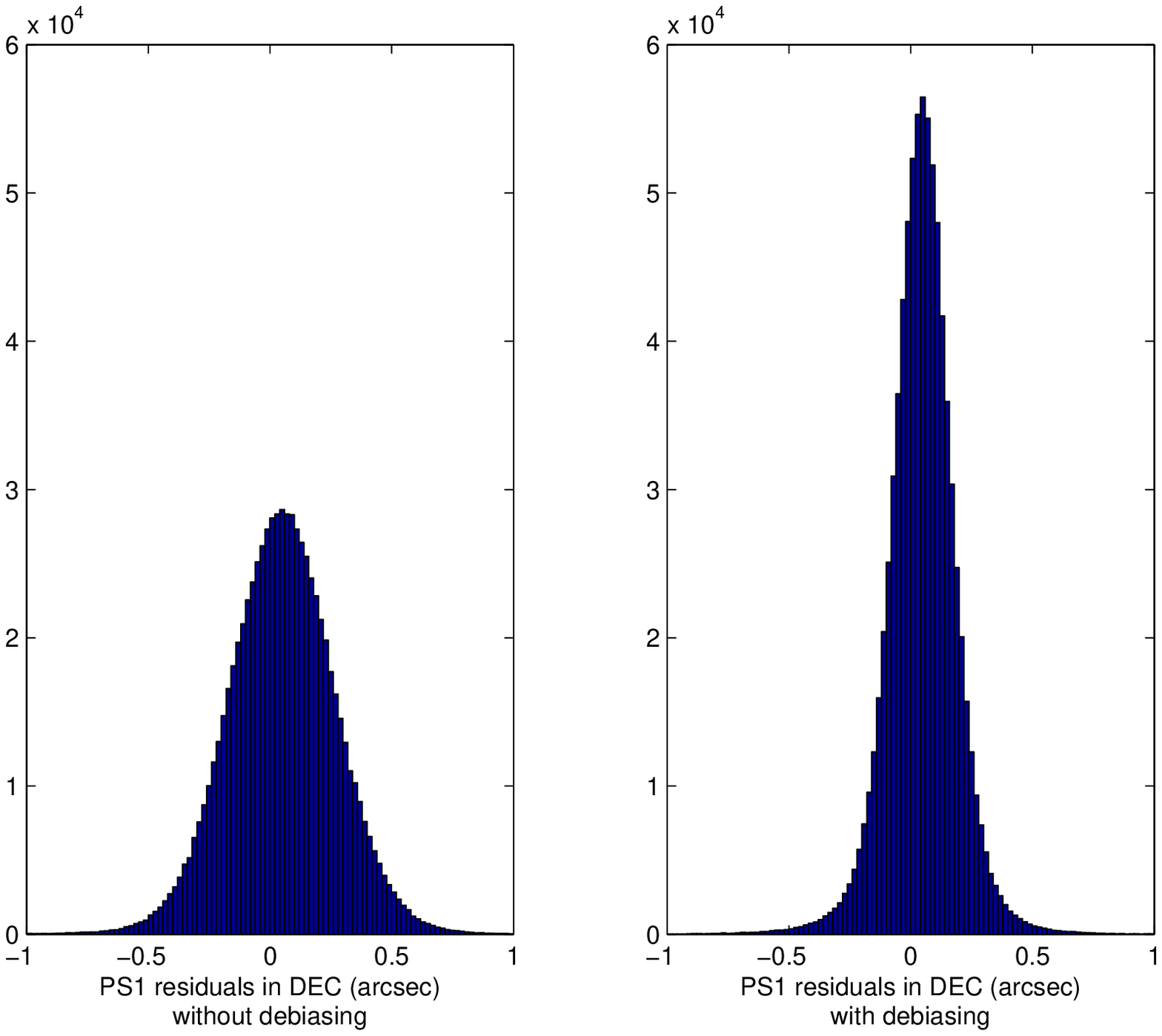}}\caption{\small
    Histograms of declination residuals for PS1 observations of
    numbered asteroids that were submitted to the MPC through 15 May
    2011. Left: with respect to orbits computed with the biased star
    catalog. Right: after applying debiasing according to
    \citet{cbm10}.}
  \label{fig:rd_F51}
\end{figure}

Another way to quantify the improvement of the orbit fit is to measure
the residuals with respect to PS1 observations of numbered
asteroids. The histograms in Figure~\ref{fig:rd_F51} show a
significant improvement in the declination residuals due to the
debiasing; the effect is also significant but less pronounced in right
ascension. The debiasing procedure improves the RMS of the residuals
from $0.21$ to $0.18$ arcsec in RA and from $0.23$ to $0.16$ arcsec in
DEC.

To take advantage of the significant improvements obtained using the
\citet{cbm10} debiasing the orbit catalogs used for all the tests
described in the next two sections have been computed in this way.
The technique is now implemented in the free software OrbFit, version
4.2 and later\footnote{http://adams.dm.unipi.it/orbfit}. The debiased
orbit catalogs are now available from the AstDyS online service.

\section{Tests with PS1 data} 
\label{s:panstarrs}

The Pan-STARRS prototype telescope, PS1, has a $1.8$~m diameter
primary (with an equivalent aperture of $1.55$~m after accounting for
the large secondary), a $1.4$~GigaPixel camera, and $7.4$ square
degree field of view \citep{hodapp2004}.  The PS1 sky survey is
intended to serve multiple scientific goals \citep[e.g., in the solar
  system][]{PS1neosymp} for which accurate astrometry is just one of
many necessary requirements. The all-sky coverage (from Hawaii)
combined with the pixel scale of $0.26$ arcsec makes PS1 potentially
capable of excellent astrometry of solar system objects.

The PS1 survey began acquiring engineering data in 2009 and the first
tests with our KNOWN\_SERVER software that implements the algorithms
described herein were run on data taken in June/July of that
year. These tests showed that the PS1 astrometry of small solar system
objects was extremely good with a standard deviation of $\sim 0.12$
arcsec in right ascension, $\sim 0.13$ arcsec in declination, and a
very small bias.  (The small bias was due to chance as the observed
region of the sky had less star catalog errors than average as
discussed below.)

PS1 officially began surveying in May 2010 but the first year's data
set is not fully homogeneous because there were many adjustments to
the telescope, camera, survey scheduling, image processing,
etc. Nevertheless, the full set of PS1 observations allows us to
assess the way the biases change with right ascension. We achieve a
more detailed analysis of the PS1 system performance using a smaller
but homogeneous dataset.

\subsection{Biases in PS1 residuals}

We can assess the regional residual systematic errors in the star
catalogs using PS1 attributions to numbered asteroids observed through
April 2011 that are widely distributed on the celestial sphere.  Note that
no debiasing is applied for the PS1 astrometry since it already uses
the 2MASS catalog \citep{2MASS} that was used as the reference catalog
in the debiasing procedure \citep{cbm10}.

\begin{figure}
\centering
\includegraphics[width=10cm]{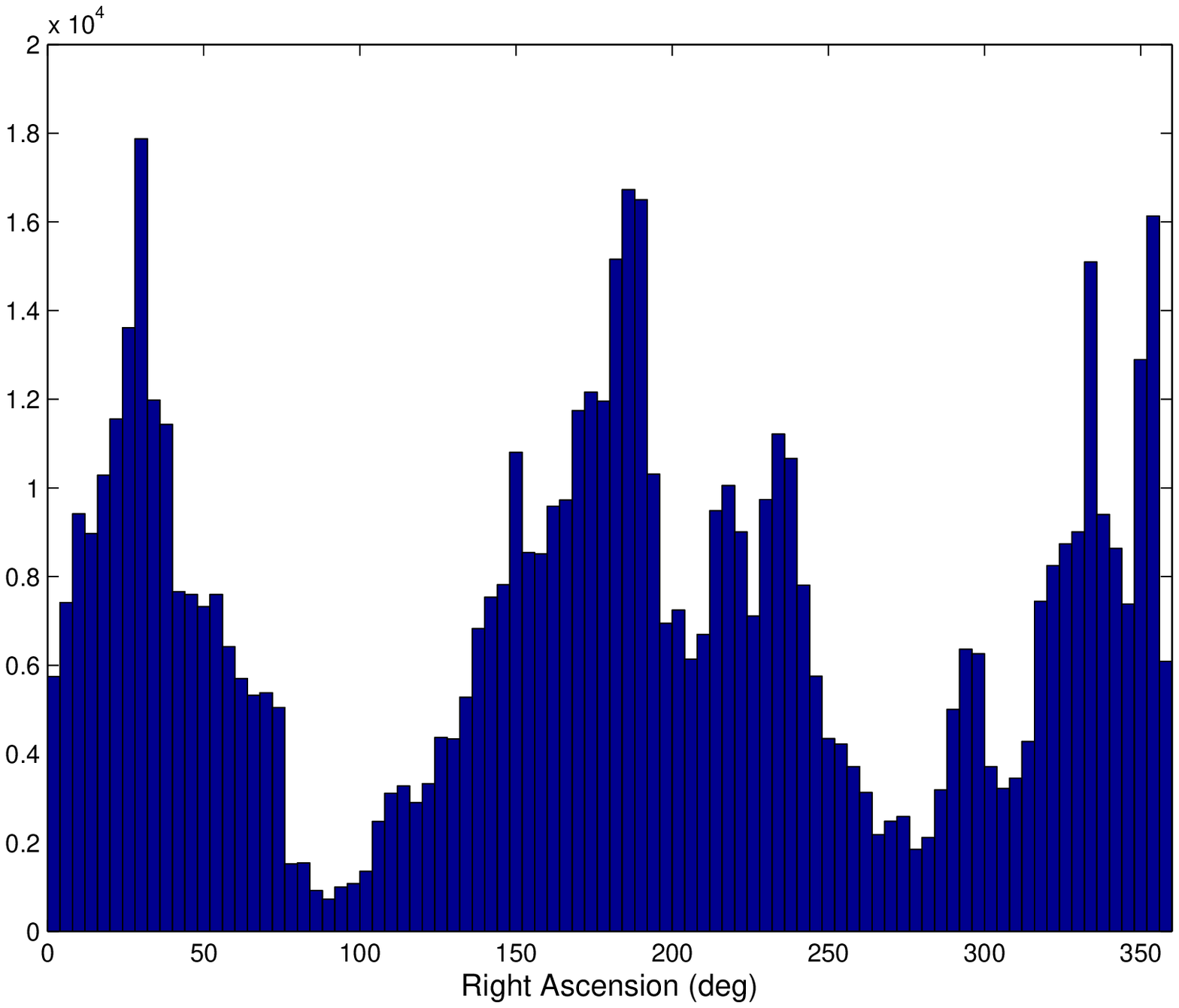}
\includegraphics[width=10cm]{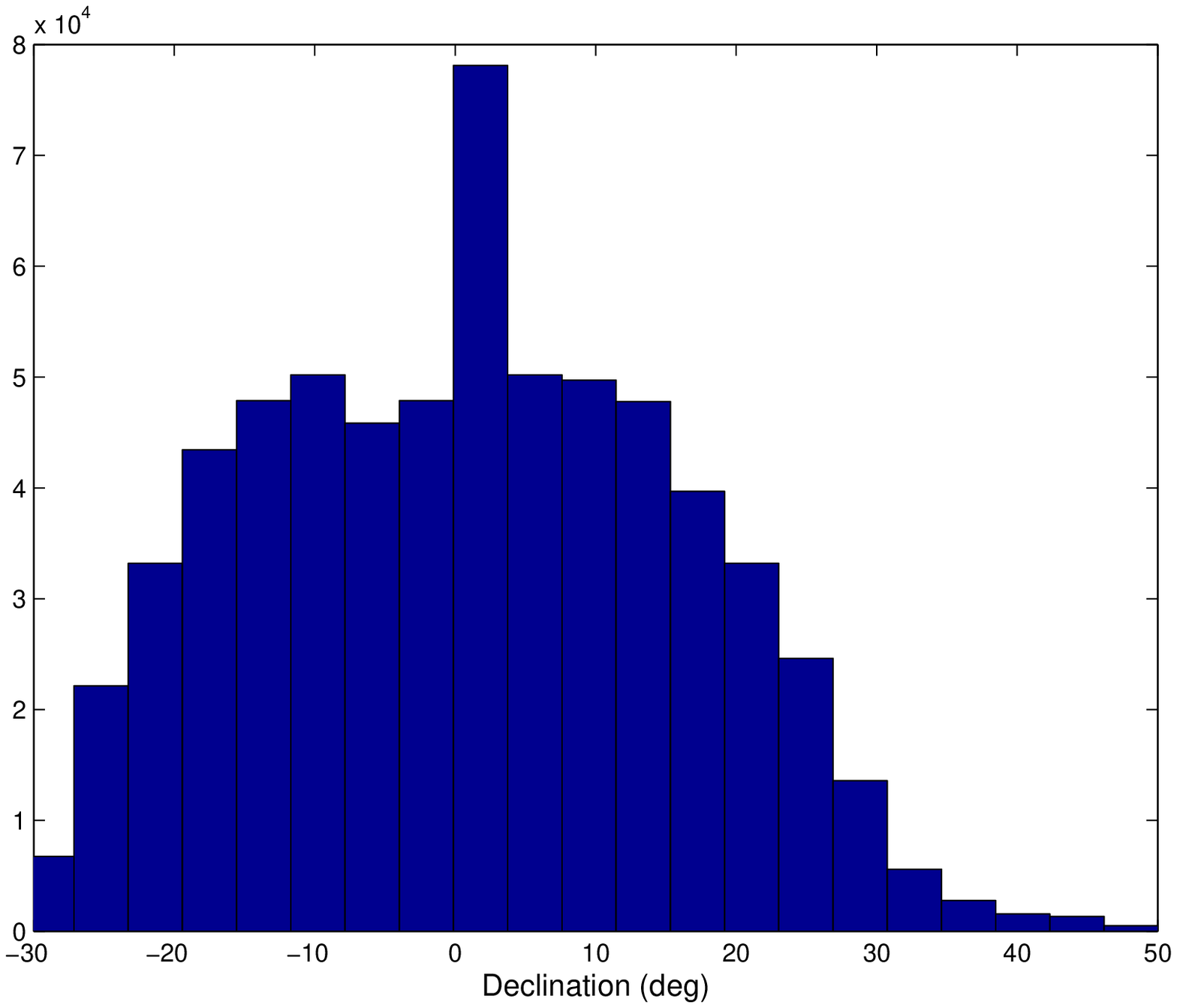}
\caption{(Top) Right ascensions and (bottom) declinations of PS1
  observations attributed to numbered asteroids up to April 2011.  The
  peak in the bin just above $0^\circ$ declination is due to the
  Medium Deep fields (see Section~\ref{s:rmscorr}).}
\label{fig:RA_DEC_hist}
\end{figure}

Figure~\ref{fig:RA_DEC_hist} shows the number distribution of the
observations as a function of right ascension and declination.  The
dramatic dips in the RA number distribution at about $105^\circ$ and
$285^\circ$ are where the galactic plane crosses the equator. The PS1
system has a reduced asteroid detection efficiency in the galactic
plane because of stellar over-crowding. The distribution is also not
uniform in declination but there are enough data for our purposes over
most of the range between $-30^\circ$ and $+50^\circ$.

\begin{figure}
\centering
\includegraphics[width=10cm]{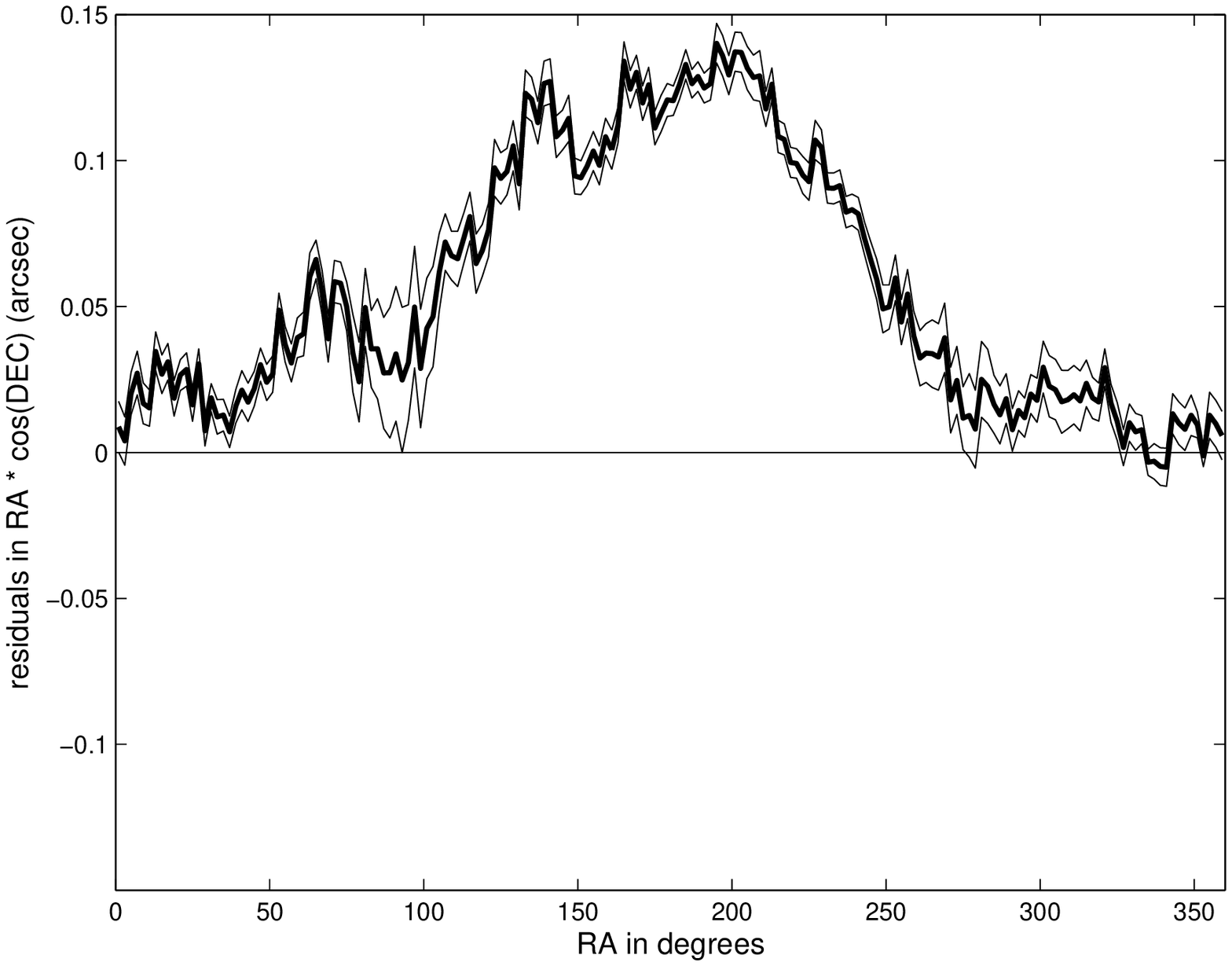}
\includegraphics[width=10cm]{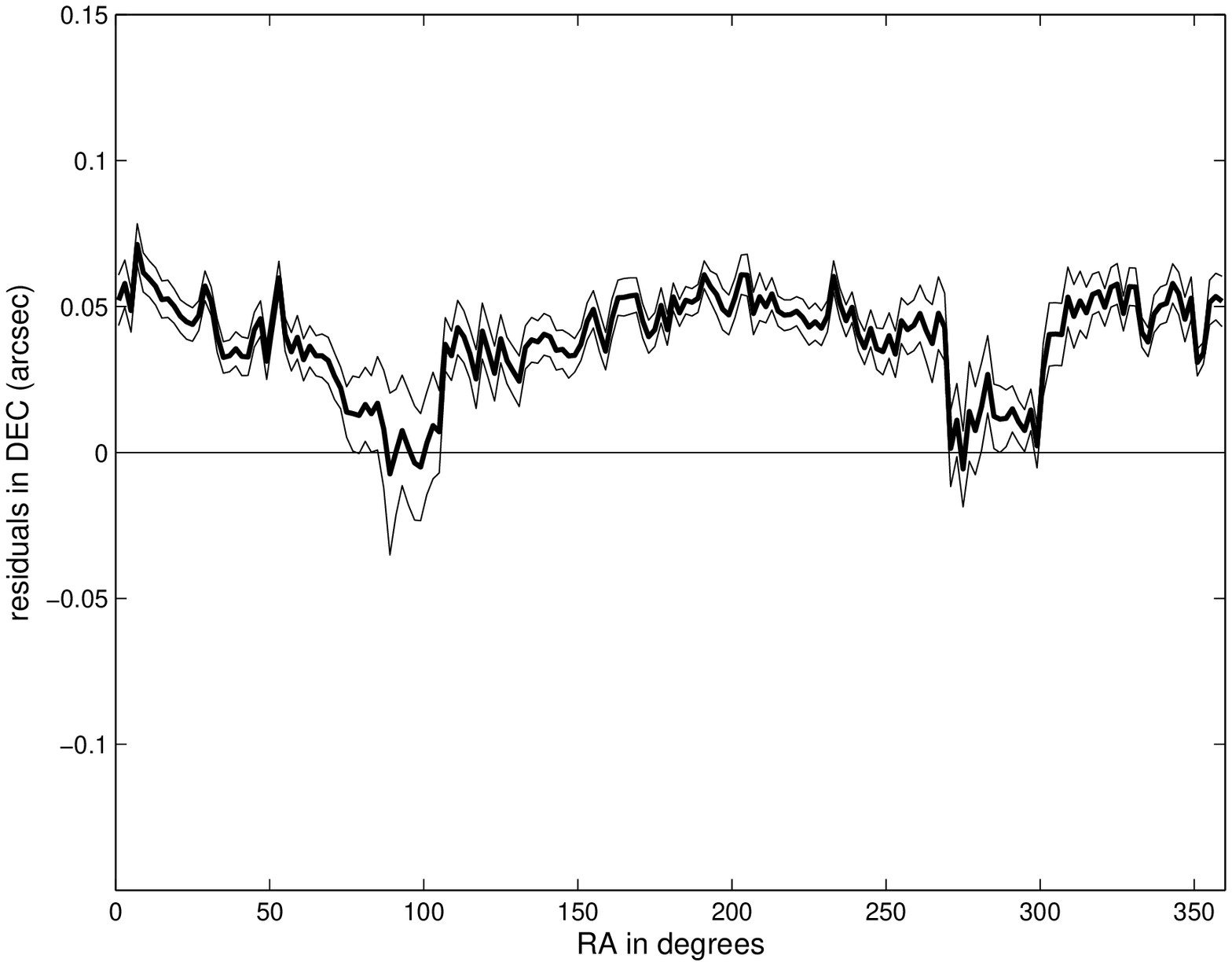}
\caption{(Top) The middle (thick) curve represents the mean residual
  in right ascension (multiplied by $\cos\delta$) as a function of
  right ascension over 2 degree wide bins.  The lower and upper curves
  correspond to $\pm3$ standard errors on the mean.  (Bottom) The
  same for the declination residual.}
\label{fig:RAresRA_DEC}
\end{figure}

The right ascension residual bias is strongly dependent on right
ascension as shown in the top panel of Figure~\ref{fig:RAresRA_DEC}.
There is a pronounced maximum of more than $0.1$ arcsec between
$110^\circ$ and $220^\circ$. Note that our first test data set of
June/July 2009 was taken between $270^\circ$ and $300^\circ$ of right
ascension where the biases appear to be only around $0.02$ arcsec. On
the contrary, there is only a weak dependence of the declination
residual biases upon right ascension.

\begin{figure}
\centering
\includegraphics[width=10cm]{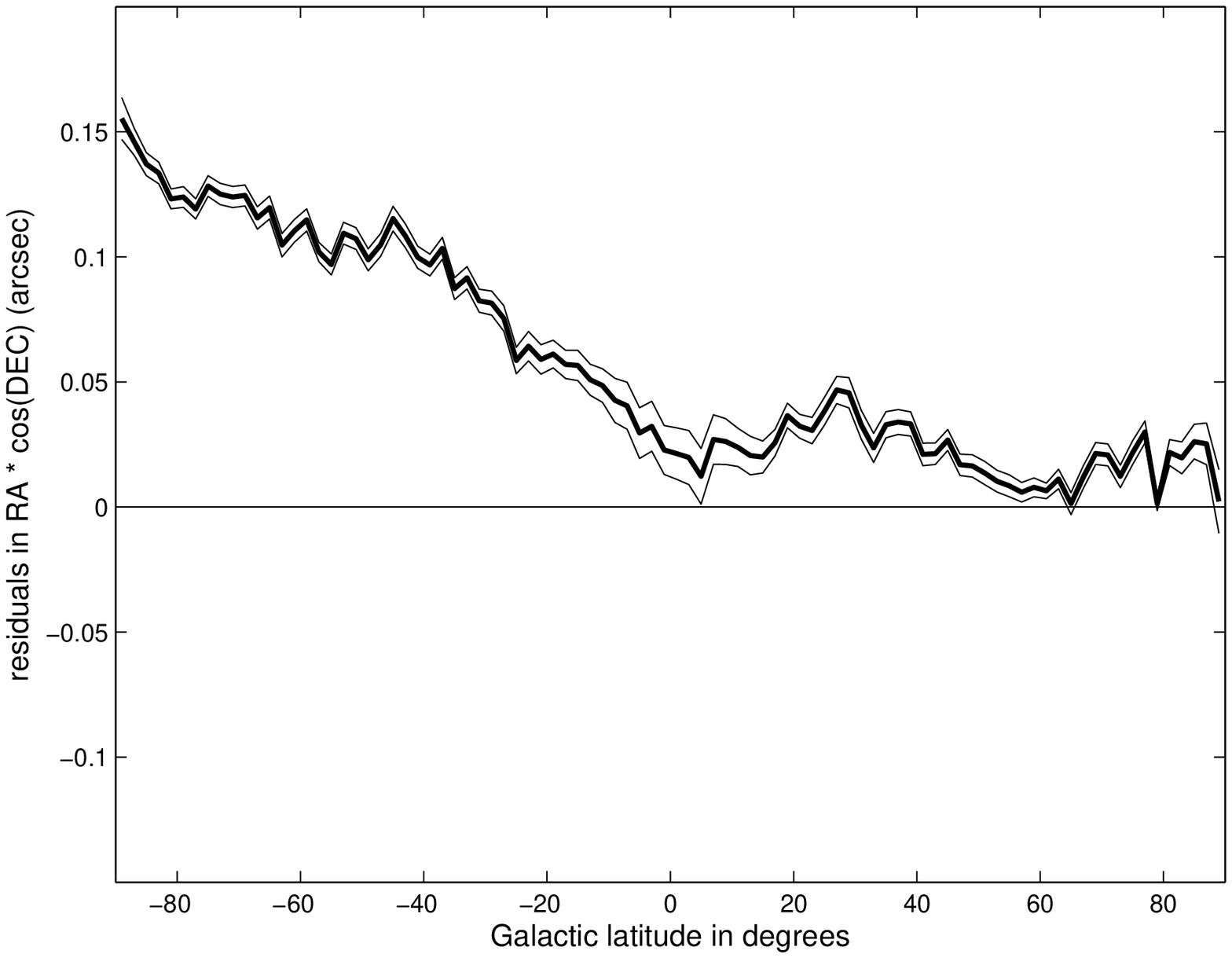}
\includegraphics[width=10cm]{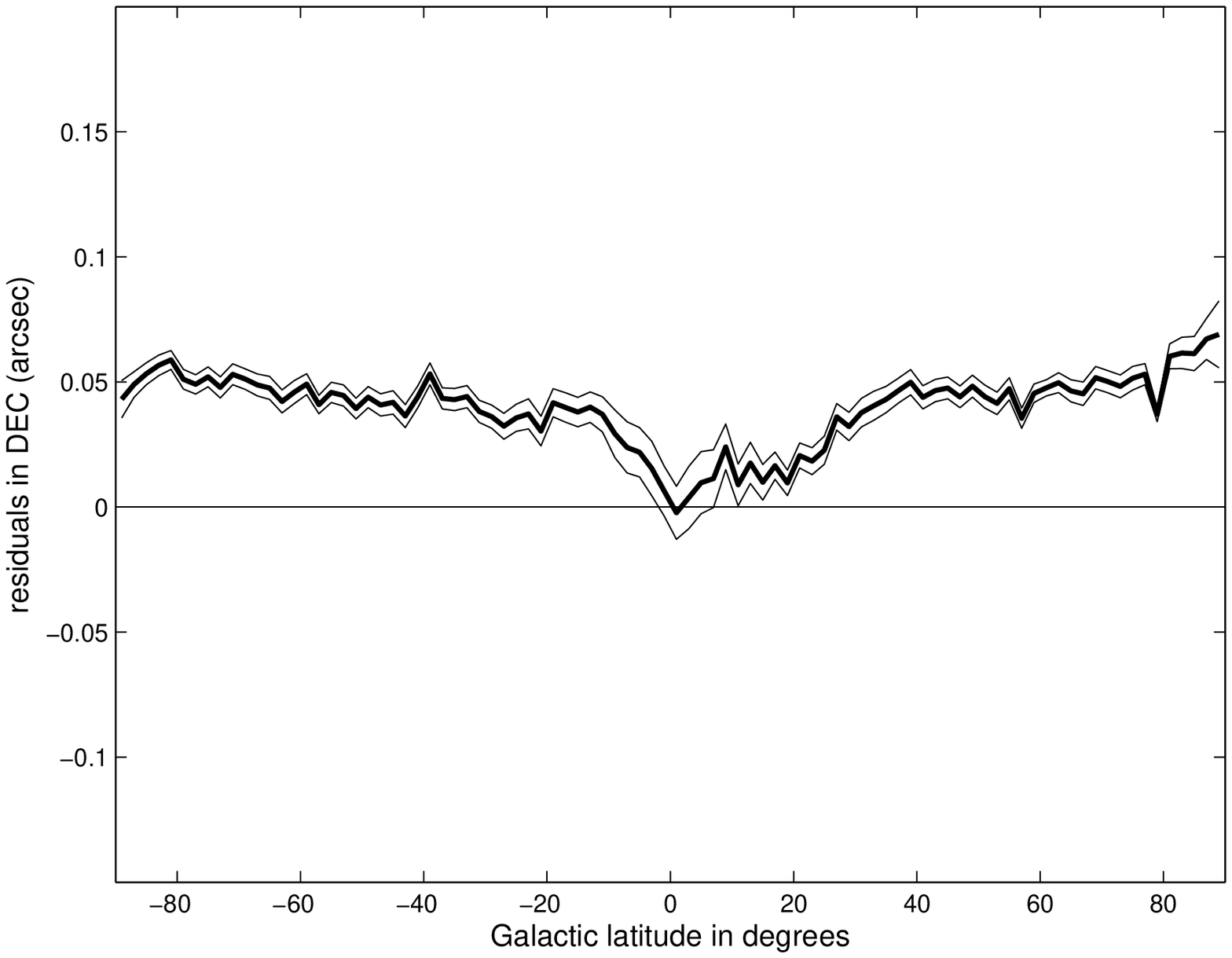}
\caption{(Top) The middle (thick) curve represents the mean residual
  in right ascension (multiplied by $\cos\delta$) as a function of
  galactic latitude over 2 degree wide bins.  The lower and upper
  curves correspond to $\pm3$ standard errors on the mean.  (Bottom)
  The same for the declination residual.}
\label{fig:RA_DECresGalLat}
\end{figure}

The location of the dips to $\sim 0.00$ arcsec for the declination
residual (see bottom panel of Figure~\ref{fig:RAresRA_DEC}) match the
ranges in RA with limited asteroid statistics as shown in
Figure~\ref{fig:RA_DEC_hist} --- where the galactic plane crosses the
equator.  Indeed, Figure~\ref{fig:RA_DECresGalLat} highlights the
relationship between the biases of the residuals and galactic
latitude.  It is absolutely clear that where the stellar sky-plane
density is high on the galactic equator the residuals disappear.

\begin{figure}
\centering 
\includegraphics[width=10cm]{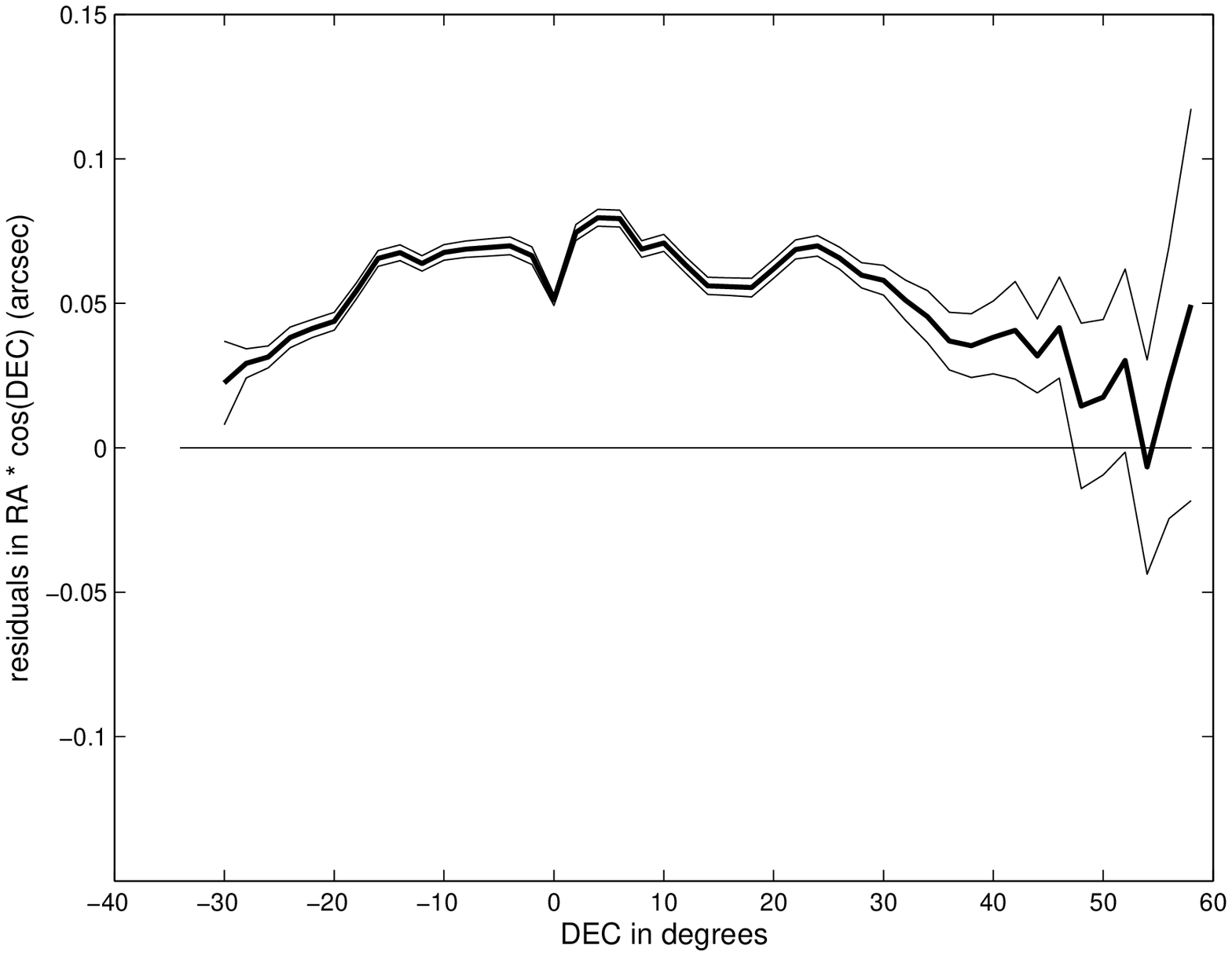}
\includegraphics[width=10cm]{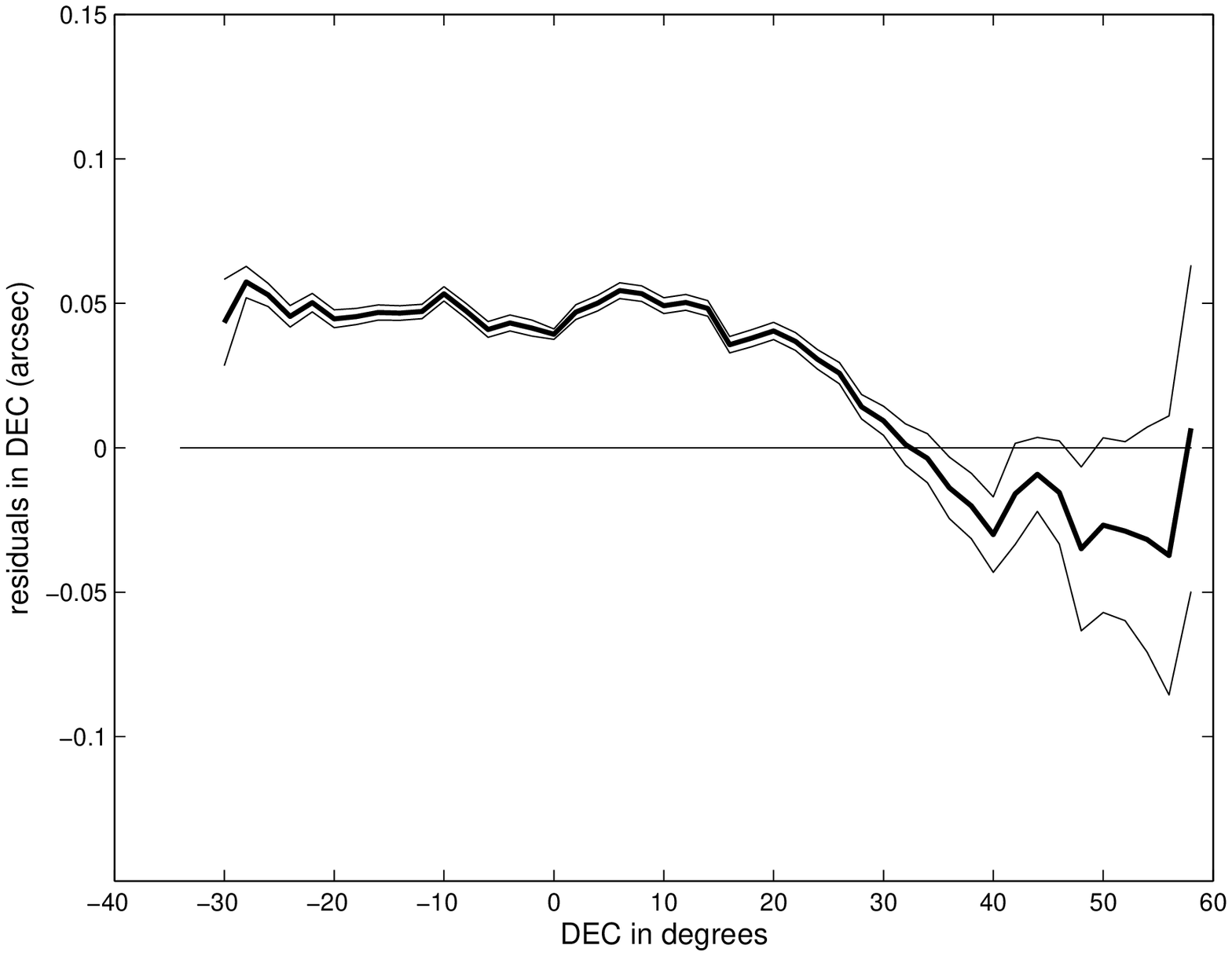} 
\caption{(Top) The middle (thick) curve represents the mean residual
  in right ascension (multiplied by $\cos\delta$) as a function of
  declination over 2 degree wide bins.  The lower and upper curves
  correspond to $\pm3$ standard errors on the mean.  (Bottom) The same
  for the declination residual.}
\label{fig:DECresRA_DEC}
\end{figure}

The dependence of right ascension residual biases upon declination
shown in the top panel of Figure~\ref{fig:DECresRA_DEC} has a
significant signature but with lower amplitude that never exceeds
$0.08$ arcsec.  Another significant effect is in the dependence of
declination residual bias upon declination as shown in the bottom
panel.  There is a relatively constant bias residual of $\sim 0.05$
arcsec from about $-30^\circ$ to $20^\circ$ declination but then a
pronounced trend to negative biases as the declination increases to
$\sim 50^\circ$.

The interpretation of these systematic regional biases would require a
dedicated study but we may have uncovered regional biases in the
2MASS star catalog that are significantly smaller than those of other
frequently used star catalogs but still relevant when used at the
level of accuracy made possible by instruments like PS1. In the medium
term we expect this problem would be removed by debiasing the 2MASS
catalog in turn with an even better catalog such as the one to be
produced by the GAIA mission.

\subsection{Standard deviations and correlations in PS1 data} 
\label{s:rmscorr}

To assess the current PS1 astrometry we have used a homogeneous
dataset processed with the same image processing software and
astrometric reduction based upon the 2MASS star catalog.  As noted
above, the \citet{cbm10} bias model uses 2MASS as the reference
catalog so that the PS1 data do not need debiasing but the historic
data from other observatories have been debiased. The attributions to
known numbered and multi-opposition objects was performed with the
KNOWN\_SERVER software with uniform configuration options such as the
quality control parameters.

The data belong to lunations 138 and 139 from 3/25/2011 to 4/17/2011
and 4/18/2011 to 5/17/2011 respectively. In each lunation there are
tracklets generated in two different types of surveys: 1) the $3\pi$
survey including fields within about $\pm 30^\circ$ of the longitude
of opposition with four exposures per night and 2) the medium deep
survey (MD) with eight exposures per visit at a few specific
bore-sights (the same bore-sight might be visited more than once on
the same night in different filters).  The $3\pi$ opposition fields
are typically acquired in the $g_{P1}$, $r_{P1}$ and $i_{P1}$ filters
and the MD surveys also use the $z_{P1}$ and $y_{P1}$ filters (see
\citet{schlafly2012} for a detailed description of the PS1 filter
system).  As mentioned above, the $3\pi$ fields cover a wide area near
opposition while the boresite locations of the 5 MD fields used in this
work are provided in Table~\ref{tab:MDfields}.

\begin{table}[h]
\centering
\caption{Identifiers, boresite locations and number of reported asteroid observations from the PS1 Medium Deep fields used in this work.}
\vspace{3mm}
\begin{tabular}{|c|c|c|c|}
  \hline
  MD Field  &  RA (deg)  &  Dec. (deg)  &  \# of obs. \\
  \hline
  \hline
  MD03  &  130.591  &  44.316  &   310  \\
  \hline
  MD04  &  150.000  &   2.200  &  5742  \\
  \hline
  MD05  &  161.916  &  58.083  &    41  \\
  \hline
  MD06  &  185.000  &  47.116  &   287  \\
  \hline
  MD07  &  213.704  &  53.083  &     6 \\
  \hline
\end{tabular}
\label{tab:MDfields}
\end{table}

\begin{table}[h]
\centering
\caption{Mean, RMS and correlation of PS1 right
  ascension and declination residuals for two lunations and two types
  of PS1 surveys (mean and RMS in arcsec). The last column is the
  number of data points in the measurement.  The rows in bold are
  the results after removing the mean (i.e., the standard deviation).}
\vspace{3mm}
\begin{tabular}{|r|l|l|l|l|l|l|r|}
  \hline
  Survey-  & \multicolumn{3}{|c|}{right ascension}& \multicolumn{3}{|c|}{declination}& number\\
  \cline{2-7}
  Lunation & Mean &  RMS &  Corr                   & Mean & RMS &  Corr               & resid.  \\
  \hline
  \hline
  $3\pi$-138 &  0.115 & 0.170 & 0.759 & 0.078 & 0.147 & 0.691 & 67566\\
             & {\bf 0.} & {\bf 0.125} & {\bf 0.542} & {\bf 0.} & {\bf 0.125} & {\bf 0.587} & \\
  \hline
  $3\pi$-139 & 0.088 & 0.154 & 0.609 & 0.073 & 0.149 & 0.614 & 44111\\
             & {\bf 0.} & {\bf 0.127} & {\bf 0.420} & {\bf 0.} & {\bf 0.129} & {\bf 0.480} & \\
  \hline
   MD-138    & 0.035 & 0.120 & 0.305 & 0.014 & 0.120 & 0.278 & 3533\\
             & {\bf 0.} & {\bf 0.115} & {\bf 0.242} & {\bf 0.} & {\bf 0.119} & {\bf 0.269} & \\
  \hline
   MD-139    & 0.038 & 0.112 & 0.332 & 0.019 & 0.109 & 0.340 & 3047\\
             & {\bf 0.} & {\bf 0.106} & {\bf 0.257} & {\bf 0.} & {\bf 0.107} & {\bf 0.323} & \\
  \hline
\end{tabular}
\label{tab:deb_panst}
\end{table}

Table~\ref{tab:deb_panst} gives the basic statistical properties of
the residuals for all four data subsets. As expected from the
discussion in the previous subsection the bias in both RA and DEC is
sensitive to the sky location and the RMS (computed with respect to
zero) changes significantly as a consequence. The RMS computed with
respect to the mean is less sensitive but it too can change within a
range of $0.02$ arcsec.

We also computed the correlations among the residuals of the same
coordinate that belong to the same tracklet. They have a large spread
that depends on the observed region in the sky and also upon the
removal of the constant bias (the mean of the residuals). This is not
a surprise since if the data are processed ignoring the presence of a
bias then the bias reappears as a correlation. The only unexpected
result is the size of this effect with apparent correlations up to
$0.76$ in RA and $0.69$ in DEC. In the $3^\circ$ wide MD field the
overall bias is much less because the target area is subject to low
biases in the 2MASS catalog and the removal of the overall mean is
almost equivalent to a direction-sensitive debiasing.  This results in
in-tracklet correlations in the range $0.24$ to $0.32$. These values
are a good measure of the intrinsic correlation of data taken over a
short time span (such as data from the same tracklet).

These results imply that we are not yet ready to build a PS1 error
model that fully accounts for biases and correlations because these
two effects cannot yet be neatly separated. To use all the information
contained in the precise PS1 observations we would need to first apply
a 2MASS catalog \citep{2MASS} debiasing, measure the correlation, and then build a
PS1-specific error model and explicitly account for the correlations
(as was done by \citet{carpino03, baer11}). Since this is a significant
undertaking it needs to be the subject of our future work and we need
to find a method to handle the data in a statistically correct manner.

\subsection{Indications for a Pan-STARRS error model}

There are two main conclusions to be drawn from the analysis of the
PS1 data.

First, the precision of the PS1 astrometry is very good. The random
component of the error model for the residuals in both RA and DEC has
a standard deviation ranging between $0.106$ and $0.129$ arcsec (see the
bold lines in Table~\ref{tab:deb_panst}). This is reasonable taking
into account the typical PS1 point spread function (PSF) full-width at
half-max of $\sim 1.1$ arcsec, the pixel scale of $\sim 0.26$ arcsec,
that asteroids move slightly during the PS1 exposures, and that the
vast majority of the reported PS1 asteroid detections have a
signal-to-noise ratio (SNR) in the range of 5-10 (i.e., for
over-sampled PSFs we expect the astrometric precision to be $\sim
FWHM/(2.4\times SNR)$, \citet{nw1995}).  Only a few professional
asteroid observers regularly achieve this astrometric accuracy and of
the large surveys only the Sloan Digital Sky Survey appears to have
reached such precision.  The PS1 advantage is that it achieves this
accuracy over the entire sky during a long term survey and will
generate a huge volume of data.  In 2010 PS1 was the fifth largest
contributor of numbered asteroid observations to the MPC with
$548,518$ (in the first 11 months of 2011 it was fourth with $935,
001$).

Second, the accuracy of the PS1 astrometry is limited by the presence
of systematic errors in the reference star catalog. Although 2MASS
\citep{2MASS} is the best possible choice for a star catalog it still
contains systematic effects at up to the level of $0.1$ arcsec that
are easily detectable with the modern generation of sky surveys. As a
consequence, it is not yet possible to fully capitalize on the
astrometric precision available from surveys like PS1. The problem
could be solved either with new and improved star catalogs (e.g., the
ones expected from the GAIA mission) or by building a survey-specific
PS1 star catalog that is referenced to a catalog with smaller regional
biases such as Tycho-2 or Hipparcos. In addition to the
position-dependent biases there is also an intrinsic correlation among
the observations belonging to the same tracklet that we are not able
to model accurately at this time.

To ensure that the PS1 observations yield the best possible nominal
solutions and reliable covariance estimates the PS1 residuals must be
weighted in a manner that accounts for systematic errors and their
correlations. That is, if we use a model that ignores biases and
correlations the confidence region of the observations of the same
tracklet needs to be a sphere containing the ellipsoid representing
the confidence region of the debiased and correlated model. If we
assume the correlated covariance matrix of the RAs of the tracklets
with $m$ observations has variances $\sigma^2$ on the diagonal and all
correlations are equal to $r$ (i.e., the covariances are all
$r\,\sigma^2$) then the largest eigenvalue is
$(1+(m-1)\,r)\cdot\sigma^2$ \citep[Section 5.8]{orbdet}. With $m=4$,
$r=0.759$, $\sigma=0.170$ the maximum eigenvalue is $0.308^2$. Thus,
the worst case from Table~\ref{tab:deb_panst} requires a weight of
$1/0.308$ arcsec$^{-1}$.  Indeed, AstDyS uses a weight of $1/0.3$ to
process the PS1 data --- the highest weight for any asteroid survey.
Some asteroid programs with either special instruments/methods or very
labor intensive reduction procedures do have higher weights.

We are not claiming that this weighting scheme is the best or only
solution for handling high precision data from modern surveys --- only
that it is a prudent way to use the PS1 data while waiting for a more
sophisticated error model that could be obtained by a better bias
removal and by an explicit correlation model.

\section{Accuracy and efficiency}
\label{s:acceff}

The purpose of this section is to measure the accuracy and the
efficiency of our asymmetric attribution procedure. \textit{Accuracy}
is the fraction of correct attributions among those proposed by our
method while \textit{efficiency} is the fraction of possible
attributions that were found.

\subsection{Accuracy test}

It is very difficult to measure accuracy with real data because it is
impossible to know the \textit{ground truth} (i.e., which attributions
are true/false) while measuring accuracy with simulated data would be
less convincing because it is difficult to simulate all possible
causes of false data (at the single detection, tracklet, and
attribution level).  In any event, false data tend to generate
tracklets with randomly oriented angular velocity that are less likely
to be attributed to a real asteroid orbit.  We therefore consider an
attribution test with real tracklets and real objects to be more
valuable at least for the same number of tracklets under
consideration.

Our solution has been to use the fact that our dataset of proposed
attributions can be split in two disjoint subsets: numbered and
multi-apparition asteroids.  The set of all attributions to numbered
objects is robust and contains only a very small fraction of false
attributions. If a tracklet has a successful attribution to a numbered
asteroid then it cannot be attributed to a different asteroid. Thus
any tracklet that is already attributed to a numbered asteroid would
almost certainly be a false attribution if it could also be attributed
to a multi-apparition asteroid.
 
In our test we used the set of $13,729$ tracklets attributed to
numbered asteroids in lunation 139 of the PS1 $3\pi$ survey and
attempted to find attributions to a list of $140,225$ multi-apparition
orbits using our algorithm for the asymmetric case. The
multi-apparition orbits provide much less accurate ephemerides and
finding an observation inside the confidence ellipse of one of the
many tested orbits is not a rare event.  The first filter provided
$44,675$ candidate attributions, the second filter reduced the number
to 8, and the third yielded 3 tracklets incorrectly attributable to
two asteroids: 2010 GU$_{104}$ (2 tracklets) and 2010 GK$_2$ (1
tracklet).

For comparison, our algorithm attributed $3,619$ tracklets to $2,950$
multi-apparition asteroids in the same lunation after removing all the
tracklets that were already attributed to numbered asteroids.  Thus,
the false attribution rate can be estimated at either $3/3,619$ or
$2/2,950$ which is $<1/1,000$ per lunation per observable object.
This result is statistically good but cannot be ignored because such
false attributions would introduce permanent `damage' in the orbit
database\footnote{\emph{False facts are highly injurious to the
    progress of science, for they often endure long...}, C. Darwin,
  \emph{The Origin of Man}, 1871.}.

The false attribution rate can be reduced to zero through additional
controls. The two asteroids 2010 GU$_{104}$ and 2010 GK$_2$ for which
false attributions were found have a historical set of observations
that only weakly constrain the orbit: two apparitions widely spaced in
time (1999--2010 and 2001--2010 respectively) and not many
observations (16 and 18 respectively). Thus, the $1\,\sigma$ ephemeris
confidence ellipses at the time of the incorrect attribution have
major semiaxes of about $47$ and $81$ arcmin, respectively. This
explains why it was possible to find a false attribution passing the
quality controls in a sample of $>13,700$ tracklets.  Moreover, in the
first case the two attributed tracklets were from a single night and
in the second there was only one tracklet.  Thus, another filter
should not accept single-night attributions to objects that have been
observed in only two previous apparitions. As a matter of fact, the
Minor Planet Center has enforced this rule for a long time and we
acknowledge that this rule is a meaningful caution to avoid
contamination of the orbits database by spurious attributions such as
the ones found in our test.

\subsection{Efficiency test}

Determining the efficiency of our attribution method suffers from the
same concerns addressed above regarding the use of real or synthetic
data.  Once again, we decided to rely on real data and used the AstDyS
catalog of numbered and multi-apparition asteroids to identify a set
of asteroids whose ephemeris definitely places them in each field of
view.  The PS1 limiting magnitude is {\it fainter} than most known
asteroids when they are at opposition so we used objects identified in
one night of morning \textit{sweet spot} observations (looking
eastwards before sunrise at solar elongations between $60^\circ$ and
$90^\circ$ in the $w$ filter).  Main belt asteroids are fainter in the
sweet spots than at opposition due to distance and phase angle effects
so that their range in apparent magnitude is better suited to
determining the PS1 sensitivity as a function of the object’s
brightness.

Then we ran KNOWN\_SERVER using the same catalog of orbits and all the
observed tracklets in the same fields to identify objects in the list
of those which could have been observed because they were in the field
of view of the PS1 sensor.  The efficiency is simply the fraction of
observable objects that were actually detected.  If the false
attribution rate is $<1/1,000$ as discussed in the previous subsection
the statistics of the successful attributions will not be
significantly contaminated.

Figure~\ref{fig:known5863} shows that the attribution efficiency as a
function of predicted apparent \textit{V} magnitude has a sharp
decline at the limiting magnitude of $\sim 21$ (where the efficiency
drops to half the peak value). The peak efficiency of
$\epsilon_{max}=0.78\pm 0.04$ occurs at $V \gtrsim 19$ but decreases
by less than a standard deviation in the brightest bin near $V\sim 18$
to $0.74\pm 0.04$.  Note that the well known offset of a few tenths of
a magnitude between predicted and actual asteroid $V$ magnitudes
\citep{juric2002} is unimportant --- all we care about here is the
peak efficiency independent of magnitude.

\begin{figure}[t]
\figfigeps{8 cm}{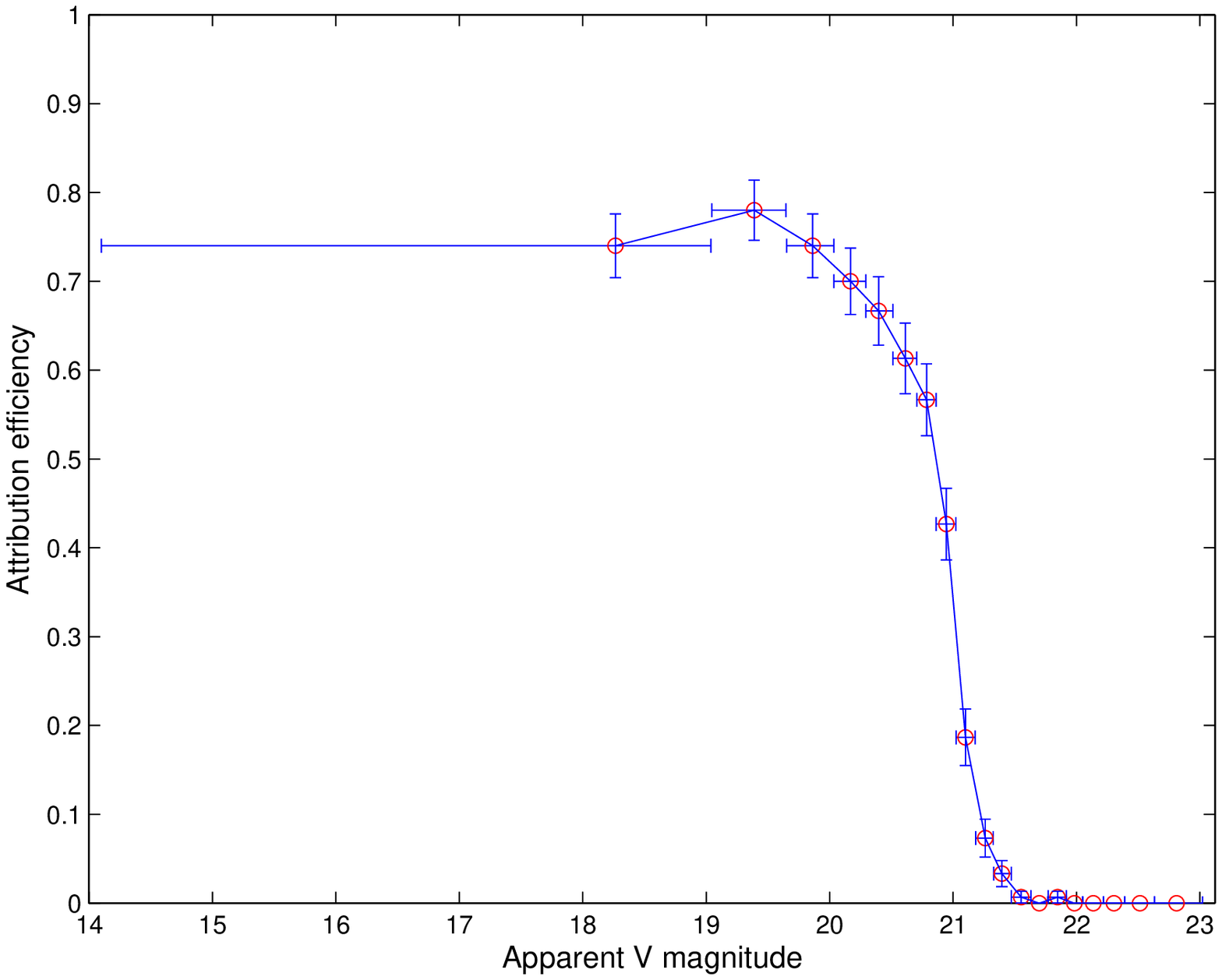}{Tracklet attribution efficiency to well
  known asteroid orbits as a function of predicted apparent \textit{V}
  magnitude from one night of the PS1 survey. Each data point
  represents 150 observable objects.  The horizontal bars indicate the
  range of values within the bin and the vertical bars represent one
  standard deviation of the estimated efficiency in the bin.}
\end{figure}

The problem in interpreting these results is that they measure several
different contributions to the efficiency including the \textit{fill
  factor} ($f$, the fraction of the focal plane covered by active
sensing pixels), the detector sensitivity ($\epsilon_D$), the
efficiency of the image processing pipeline (IPP) in detecting moving
objects ($\epsilon_{IPP}$), the Moving Object Processing System's
\citep[MOPS,\ ][]{kubica2007} efficiency at linking detections into
tracklets ($\epsilon_{MOPS}$), and the efficiency of the attribution
algorithm ($\epsilon_{attrib}$).  Disentangling each effect as a
function of $V$ is difficult but unnecessary for our purposes.
Instead, we know that the generic efficiency $\epsilon_x \ge
\epsilon_{max}$ and attempt to establish a minimum value of
$\epsilon_{attrib}$.

The PS1 camera consists of 60 orthogonal transfer array (OTA) CCDs
arranged in an $8\times 8$ array (the four corners do not have CCDs)
and each OTA contains an on-chip $8\times 8$ mosaic of `cells'.  The
camera fill factor includes the physical gaps between OTAs, smaller
gaps between cells on a single OTA, area lost to defective cells and
bad pixels, and the overlap between the sensor and the Field Of View
(FOV) of the optical system.  There are additional losses of the order
of $1\%$ specific to individual exposures due to a `dynamic mask'
applied to remove bright diffraction spikes and internal reflections.

The PS1 image processing team produces periodic analysis of sensor
fill factor as the camera tuning improves.  The most recent study (May
2010) yields $f=0.79$ due to a $7.0\%$ loss due to inter-OTA gaps,
a $4.3\%$ loss due to inter-cell gaps, and $9.7\%$ mask fraction in
the UN-vignetted 3.0-degree FOV. For our KNOWN\_SERVER efficiency
study we only consider asteroids that should appear in the 3.0-degree
FOV so $f=0.79$ is applicable.

The fill factor estimate implies that the peak efficiency (at
magnitudes between 19 and 20) leaves little room for losses due to the
IPP, MOPS, and attribution software. Each of the processing steps has
a minimum efficiency of $0.98^{+0.02}_{-0.05}$.  I.e., the overall
efficiency is dominated by the fill factor and the efficiency of the
other steps including attribution is consistent with 100\%.  The
apparent drop in the efficiency for brighter apparent magnitudes is
not significant.  It might indicate some problems in the image
processing and accurate astrometric reduction of bright detections.
The PS1 CCDs saturate at about $w=15.9$ and the $w$-filter is used for
sweet spot observations.


\section{Conclusions and future work}
\label{s:conclusions}

We proposed a new procedure to identify known asteroids among the
new observations from a survey. This procedure solves the problems due
to the asymmetry in quantity and quality between the data and historic
observations.

We tested the procedure with real data from the Pan-STARRS PS1
survey and assessed the performance of our algorithms and the
astrometric accuracy of the survey data. The main results are the
following.

First, the new algorithms are accurate and efficient. For
multi-apparition asteroids the false attribution fraction is less than
$1/1,000$ and even those can be eliminated by following the MPC's good
practice of requiring two nights of data for a recovery at a new
apparition. The algorithm's attribution efficiency is high and
consistent with 100\% but cannot really be measured because it is
entangled with other efficiency losses such as the fill factor.

Second, the PS1 data have significantly lower astrometric error than
other asteroid surveys. This error can be identified only after
removing the biases due to systematic errors in the star
catalogs. Indeed, we arrived at the conclusion that even the 2MASS star
catalog contains enough biases to affect the PS1 error model.  It is
likely that the use of other catalogs, or at least debiasing with
respect to them, would further improve the PS1 error model; e.g.,
Tycho-2 could be used now and the GAIA catalog in the future.

Third, we consider that the PS1 data can be included in the fit for
asteroid orbits with weights corresponding to $1/0.3$ arcsec$^{-1}$
that implicitly account for the effects of correlations. A model with
debiasing of the 2MASS astrometry and explicitly taking into account
the correlations could allow a further improvement in the diagonal
elements of the weighting matrix by another factor $2\sim 3$.

\subsection{Future work}

Apart from the improvements mentioned above there are three areas where
there is room for future developments.

The first is in the use of KNOWN\_SERVER, or similar algorithms and
software, as a filter to remove observations of known objects from a
new set of data to leave subset with a larger fraction of potential
new discoveries.  Although we think that this principle is valid we
have not yet been able to test the idea on real data. For the removal
of the known objects to be a significant contribution in decreasing
the false identification rate, the false tracklet rate must be below
some threshold and the number of known objects must be large compared
to the number of unknown objects at the system's limiting
magnitude. For the current PS1 data the false discoveries due to
spurious detections (that form false tracklets) are more important
than the false discoveries due to incorrect linking of true tracklets.

The second is the possibility of using KNOWN\_SERVER as an alarm to
detect unusual phenomena in well known asteroids. The most interesting
cases could be the Main Belt Comets \citep[MBC,\ ][]{hsieh2006}.

As an example, the numbered asteroid (300163) was found to be a MBC on
the basis of PS1 observations that showed an image wider than the PSF
of nearby stars \citep{pscomet}.  However, the standard KNOWN\_SERVER
output automatically flags these observations as very unusual in at
least 3 different ways.  Two of the strange flags are astrometric in
nature: $BIAS_\alpha=4.01$, corresponding to a systematic shift by
$0.85$ arcsec backward (with respect to orbital motion), and
$BIAS_\delta=2.11$, corresponding to a shift by $0.44$ arcsec
North. These could be interpreted as an effect of displacement of the
center of light with respect to the center of mass, and/or as an
effect of non-gravitational perturbations.  The other strange flag was
that the observation's apparent magnitudes were on average brighter
than the predicted ones by about $1.05$ which is also likely a
consequence of the outburst.  

The problem is to define a filter selecting anomalous behaviors such
as the one above as an indication of possible orbital, luminosity
and/or image shape changes. The main challenge is to design such a
filter with a low false positive rate that allows for dedicated follow
up of the MBC candidates.

The third is the possibility of improving the quality of other
existing large datasets (such as those generated by other sky surveys
or even the MPC asteroid astrometry database) by either detecting
missing attribution or proposing deletion of dubious ones. This work
would require the collaboration of the data providers for their
insight on the datasets' problems.

\section*{Acknowledgments}

The PS1 Surveys have been made possible through contributions of the
Institute for Astronomy, the University of Hawaii, the Pan-STARRS
Project Office, the Max-Planck Society and its participating
institutes, the Max Planck Institute for Astronomy, Heidelberg and the
Max Planck Institute for Extraterrestrial Physics, Garching, The Johns
Hopkins University, Durham University, the University of Edinburgh,
Queen's University Belfast, the Harvard-Smithsonian Center for
Astrophysics, and the Las Cumbres Observatory Global Telescope
Network, Incorporated, the National Central University of Taiwan, and
the National Aeronautics and Space Administration under Grant
No. NNX08AR22G issued through the Planetary Science Division of the
NASA Science Mission Directorate.

Some authors have also been supported for this research by: the
Italian Space Agency, under the contract ASI/INAF I/015/07/0, (A.M.,
F.B., D.F.); the Ministry of Education and Science of Serbia, under
the project 176011 (Z.K.).

The authors wish to thank the referees (T. Spahr and an anonymous one)
for their constructive comments.

\bibliographystyle{elsarticle-harv}

\begin{thebibliography}{00}

\bibitem[Baer et al.(2011)]{baer11} Baer, J., Chesley, S.~R., Milani,
  A.\ 2011.\ Development of an observational error model.\ Icarus 212,
  438-447.

\bibitem[Chesley et al.(2010)]{cbm10} Chesley, S.~R., Baer, J., Monet,
  D.~G.\ 2010.\ Treatment of star catalog biases in asteroid
  astrometric observations.\ Icarus 210, 158-181.

\bibitem[Carpino et al.(2003)]{carpino03} Carpino, M., Milani, A.,
  Chesley, S.~R.\ 2003.\ Error statistics of asteroid optical
  astrometric observations.\ Icarus 166, 248-270.

\bibitem[Granvik and Muinonen(2008)]{gra08} Granvik, M., Muinonen, K.\
  2008.\ Asteroid identification over apparitions.\ Icarus 198,
  130-137.

\bibitem[Hsieh and Jewitt(2006)]{hsieh2006} Hsieh, H.~H., Jewitt, D.\
  2006.\ A Population of Comets in the Main Asteroid Belt.\ Science
  312, 561-563.

\bibitem[Hergenrother(2011)]{pscomet} Hergenrother, C.~W.\ 2011,
  Central Bureau Electronic Telegrams, 2920, 2
  
\bibitem[Hodapp et al.(2004)]{hodapp2004} Hodapp, K.~W. et al. 2004.\
  Design of the Pan-STARRS telescopes.\ Astronomische Nachrichten 325,
  636-642.

 \bibitem[Jedicke et al.(2007)]{PS1neosymp} Jedicke, R., Magnier, 
E.~A., Kaiser, N., Chambers, K.~C.\ 2007.\ The next decade of Solar System 
discovery with Pan-STARRS.\ IAU Symposium 236, 341-352.

\bibitem[Juri{\'c} et al.(2002)]{juric2002} Juri{\'c}, M., and 15
  colleagues 2002.\ Comparison of Positions and Magnitudes of
  Asteroids Observed in the Sloan Digital Sky Survey with Those
  Predicted for Known Asteroids.\ The Astronomical Journal 124,
  1776-1787.

\bibitem[Kubica et al.(2007)]{kubica2007} Kubica, J. et al.\ 2007.\
  Efficient intra- and inter-night linking of asteroid detections
  using kd-trees.\ Icarus 189, 151-168.

\bibitem[Marsden(1985)]{marsden85} Marsden, B.~G.\ 1985.\ Initial
  orbit determination - The pragmatist's point of view.\ The
  Astronomical Journal 90, 1541-1547.

\bibitem[Milani(1999)]{ident1} Milani, A.\ 1999.\ The Asteroid
  Identification Problem. I. Recovery of Lost Asteroids.\ Icarus 137,
  269-292.

\bibitem[Milani et al.(2001)]{ident4} Milani, A., Sansaturio, M.~E.,
  Chesley, S.~R.\ 2001.\ The Asteroid Identification Problem IV:
  Attributions.\ Icarus 151, 150-159.

\bibitem[Milani et al.(2005)]{ons2} Milani, A., Gronchi, G.~F., Kne{\v
    z}evi{\'c}, Z., Sansaturio, M.~E., Arratia, O.\ 2005.\ Orbit
  determination with very short arcs. II. Identifications.\ Icarus
  179, 350-374.

\bibitem[Milani et al.(2008)]{ons3} Milani, A., Gronchi, G.~F.,
  Farnocchia, D., Kne{\v z}evi{\'c}, Z., Jedicke, R., Denneau, L.,
  Pierfederici, F.\ 2008.\ Topocentric orbit determination: Algorithms
  for the next generation surveys.\ Icarus 195, 474-492.

\bibitem[Milani and Gronchi(2010)]{orbdet} Milani, A., Gronchi, G.~F.\
  2010.\ Theory of Orbital Determination.\ Theory of Orbital
  Determination, by Andrea Milani and Giovanni F.~Gronchi.~ISBN
  978-0-521-87389-5.~Published by Cambridge University Press,
  Cambridge, UK, 2010.

\bibitem[Neuschaefer and Windhorst(1995)]{nw1995} Neuschaefer, L.~W.,
  Windhorst, R.~A.\ 1995.\ Observation and reduction methods of deep
  Palomar 200 inch 4-Shooter mosaics.\ The Astrophysical Journal
  Supplement Series 96, 371-399.
  
\bibitem[Sansaturio and Arratia(2011)]{san11} Sansaturio, M.~E.,
  Arratia, O.\ 2011.\ Mining knowledge in One Night Stands data sets.\
  Monthly Notices of the Royal Astronomical Society 1935.

\bibitem[Schlafly et al.(2012)]{schlafly2012} Schlafly, E.~F.,
  D. P. Finkbeiner, D.~P., Juri´c, M., Magnier,
  E.~A.\ 2012.\ Photometric calibration of the first 1.5 years of the
  PAN-STARRS1 survey (submitted).

\bibitem[Skrutskie et al.(2006)]{2MASS} Skrutskie, M.~F. et al. 2006.\
  The Two Micron All Sky Survey (2MASS).\ The Astronomical Journal
  131, 1163-1183.
\end{thebibliography}

\end{document}